\documentstyle[aps,epsfig]{revtex}
\begin{document}

\title{Folding model analysis of alpha radioactivity }

\author{D.N. Basu\thanks{E-mail:dnb@veccal.ernet.in}}
\address{Variable  Energy  Cyclotron  Centre,  1/AF Bidhan Nagar,
Kolkata 700 064, India}
\date{\today }
\maketitle
\begin{abstract}

      Radioactive decay of nuclei via emission of $\alpha$ particles has been studied theoretically in the framework of a superasymmetric fission model using the double folding (DF) procedure for obtaining the  $\alpha$-nucleus interaction potential. The DF nuclear potential has been obtained by  folding in the density distribution functions of the $\alpha$ nucleus and the daughter nucleus with a realistic effective interaction. The M3Y effective interaction has been used for calculating the nuclear interaction potential which has been supplemented by a zero-range pseudo-potential for exchange along with the density dependence. The nuclear microscopic $\alpha$-nucleus potential thus obtained has been used along with the Coulomb interaction potential to calculate the action integral within the WKB approximation. This subsequently yields calculations for the half lives of $\alpha$ decays of nuclei. The density dependence and the exchange effects have not been found to be very significant. These calculations provide reasonable estimates for the lifetimes of $\alpha$ radioactivity of nuclei.

\end{abstract}



\section{INTRODUCTION}
\label{sectoin1}

      One of the first successes of quantum mechanics was the explanation of $\alpha$-particle tunneling through barrier. About 20 years after the first experimental observation of the $\alpha$ radioactivity \cite{r1,r2}, the theoretical explanation in terms of quantum mechanical barrier penetration \cite{r3,r4} was provided which was, at least, qualitatively successful. Since then the $\alpha$ decay half life measurements with constantly improving techniques and the quest for theoretical explanation of their absolute magnitudes have continued unabated. The attractive features of $\alpha$ radiactivity are that the $\alpha$ particles can be detected rather easily under favourable conditions such as high efficiency, low background and good energy resolution.   

      About ten years ago, theoretical estimates of lifetimes for the $\alpha$ and various exotic decays have been provided in detail by the analytical superasymmetric fission model (ASAFM) \cite{r5,r6} with reasonable success. This was followed by the cluster model (CM) calculations for the $\alpha$  decay half lives \cite{r7} for even more comprehensive database with similar success. But the nuclear interaction potentials used by both the theoretical approaches described above are phenomenological in nature and designed to fit the experimental data. The ASAFM uses a parabolic potential approximation for the nuclear interaction potential, which is a rather unusual $\alpha$-nucleus interaction potential. The CM uses a strange cos-hyperbolic form for the nuclear interaction potential  \cite{r7}. Nevertheless, they provide a benchmark against which more microscopically based treatments can be judged.

      In the present work, the nuclear potentials have been obtained microscopically by double folding the $\alpha$ and daughter nuclei density distributions with a realistic M3Y effective interaction. This is the ideal procedure of obtaining nuclear interaction energy for the $\alpha$-nucleus interaction. Any liquid drop like properties such as surface energy are basically macroscopic manifestation of microscopic phenomena. A double folding potential obtained using M3Y effective interaction is more appropriate because of its microscopic nature. A potential energy surface is inherently embedded in this description. Moreover, the use of global microscopic nuclear potentials for a wide range of $\alpha$-nucleus interaction is also theoretically a very thorough approach. The double folding potential has then been utilised within the superasymmetric fission model description. This superasymmetric fission model using microscopic potentials has been used to provide estimates of $\alpha$ decay half lives for a wide range of $\alpha$-emitters. The semirealistic explicit density dependence on the M3Y effective interaction has also been employed to study the effects of density dependence on the $\alpha$ decay lifetimes. The penetrability of the pre-scission part of the potential barrier provides the $\alpha$ cluster preformation probability within a superasymmetric fission model \cite{r6} .  

\section{FORMALISM}
\label{section2}

      The microscopic nuclear potentials $V_N(R)$ have been obtained by double folding in the densities of the fragments $\alpha$ and daughter nuclei with the finite range realistic M3Y effective interacion as

\begin{equation}
 V_N(R) = \int \int \rho_1(\vec{r_1}) \rho_2(\vec{r_2}) v[|\vec{r_2} - \vec{r_1} + \vec{R}|] d^3r_1 d^3r_2 
\label{seqn1}
\end{equation}
\noindent
where the density distribution function $\rho_1$ for the $\alpha$ particle has the Gaussian form

\begin{equation}
 \rho_1(r) = 0.4229 exp( - 0.7024 r^2)
\label{seqn2}
\end{equation}                                                                                                                                           \noindent     
whose volume integral is equal to $A_\alpha ( = 4 )$, the mass number of $\alpha$-particle. The density distribution function $\rho_2$ used for the residual cluster, the daughter nucleus, has been chosen to be of the spherically symmetric form given by

\begin{equation}
 \rho_2(r) = \rho_0 / [ 1 + exp( (r-c) / a ) ]
\label{seqn3}
\end{equation}                                                                                                                                           \noindent     
where                        
 
\begin{equation}
 c = r_\rho ( 1 - \pi^2 a^2 / 3 r_\rho^2 ), ~~    r_\rho = 1.13 A_d^{1/3}  ~~   and ~~    a = 0.54 ~ fm
\label{seqn4}
\end{equation}
\noindent
and the value of $\rho_0$ is fixed by equating the volume integral of the density distribution function to the mass number $A_d$ of the residual daughter nucleus. The distance s between any two nucleons belonging to the residual daughter nucleus and the emitted $\alpha$ nucleus is given by 

\begin{equation}
 s = |\vec{r_2} - \vec{r_1} + \vec{R}|
\label{seqn5}
\end{equation}   
\noindent
while the interaction potential between any such two nucleons $v(s)$ appearing in eqn.(1) is given by the M3Y effective interaction. The total interaction energy $E(R)$ between the $\alpha$ nucleus and the residual daughter nucleus is equal to the sum of the nuclear interaction energy, the Coulomb interaction energy and the centrifugal barrier. Thus

\begin{equation}
 E(R) = V_N(R) + V_C(R) + \hbar^2 l(l+1) / (2\mu R^2)
\label{seqn6}
\end{equation}   
\noindent
where $\mu = mA_\alpha A_d/A$  is the reduced mass, $A$ is the mass number of the parent nucleus and m is the nucleon mass measured in the units of $MeV/c^2$. Assuming spherical charge distribution (SCD) for the residual daughter nucleus and considering the $\alpha$-paticle to be a point charge, the $\alpha$-nucleus Coulomb interaction potential $V_C(R)$ is given by

\begin{eqnarray}
 V_C(R) =&&Z_\alpha Z_d e^2/ R~~~~~~~~~~~~~~~~~~~~~~~~~~~~~~~~~~for~~~~R \geq R_c \nonumber\\
            =&&(Z_\alpha Z_d e^2/ 2R_c).[ 3 - (R/R_c)^2]~~~~~~~~~~for~~~~R\leq R_c 
\label{seqn7}
\end{eqnarray}   
\noindent
where $Z_\alpha$ and $Z_d$ are the atomic numbers of the $\alpha$-particle and the daughter nucleus respectively. The touching radial separation $R_c$ between the $\alpha$-particle and the daughter nucleus is given by $R_c = c_\alpha+c_d$ where $c_\alpha$ and $c_d$ has been obtained using eqn.(4). If point charge distribution (PCD) is also assumed for the residual daughter nucleus then the $\alpha$-nucleus Coulomb interaction potential has the simple form of $V_C(R) =Z_\alpha Z_d e^2/ R$ for all R. The energetics allow spontaneous emission of $\alpha$-particles only if the released energy

\begin{equation}
 Q = M - ( M_\alpha + M_d)
\label{seqn8}
\end{equation}
\noindent
is a positive quantity, where $M$, $M_\alpha$ and $M_d$ are the atomic masses of the parent nucleus, the emitted $\alpha$-particle and the residual daughter nucleus, respectively,  expressed in the units of energy. It is important to mention here that the correctness of predictions for possible decay modes, therefore, rests on the accuracy of the ground state masses of nuclei.

      In the present work, the half life of the parent nucleus against the split into an $\alpha$ and a daughter is calculated using the WKB barrier penetration probability. The assault frequecy $\nu$ is obtained from the zero point vibration energy $E_v = (1/2)\hbar\omega = (1/2)h\nu$. The half life $T$ of the parent nucleus $(A, Z)$ against its split into an $\alpha$ $(A_\alpha, Z_\alpha)$ and a daughter $(A_d, Z_d)$  is given by

\begin{equation}
 T = [(h \ln2) / (2 E_v)] [1 + \exp(K)]
\label{seqn9}
\end{equation}
\noindent
where the action integral $K$ within the WKB approximation is given by

\begin{equation}
 K = (2/\hbar) \int_{R_a}^{R_b} {[2\mu (E(R) - E_v - Q)]}^{1/2} dR
\label{seqn10}
\end{equation}
\noindent
where $R_a$ and $R_b$ are the two turning points of the WKB action integral determined from the equations

\begin{equation}
 E(R_a)  = Q + E_v =  E(R_b)
\label{seqn11}
\end{equation} 
\noindent
For the present calculations, from a fit to a selected set of experimental data on $\alpha$ emitters the following law \cite{r8} was found for the zero point vibration energies for the $\alpha$ decays
  
\begin{eqnarray}
 E_v =&& 0.1045.Q~~~~~~~~~~for~~even(Z)-even(N)~~parent~nuclei \nonumber\\
        =&& 0.0962.Q~~~~~~~~~~for~~odd(Z)-even(N)~~parent~nuclei  \nonumber\\
        =&& 0.0907.Q~~~~~~~~~~for~~even(Z)-odd(N)~~parent~nuclei \nonumber\\
        =&&0.0767.Q~~~~~~~~~~for~~odd(Z)-odd(N)~~parent~nuclei 
\label{seqn12}
\end{eqnarray} 
\noindent
which includes the shell and pairing effects. The values of the proportionality constants of $E_v$ with $Q$ is the largest for even-even parent and the smallest for the odd-odd one. If all other conditions are the same one may observe that with greater value of $E_v$, the life time is shortened indicating higher emission rate. The shell effects of $\alpha$ radioactivity is implicitly contained in the zero point vibration energy due to its proportionality with the Q value, which is maximum when the daughter nucleus has a magic number of neutrons and protons. 

\section{CALCULATIONS}
\label{section3}

      The two turning points of the action integral given by eqn.(10) have been obtained by solving eqns.(11) using 
the microscopic double folding potential given by eqn.(1) along with the Coulomb potential given by eqn.(7) and the centrifugal barrier. Then the WKB action integral between these two turning points has been evaluated numerically using eqn.(1), eqn.(6), eqn.(7), eqn.(8) and eqn.(12). The calculations have been performed using $v(s)$, inside the integral of eqn.(1) for the double folding (DF) procedure, as only the M3Y effective \cite{r9} interaction

\begin{equation}
  v(s) = 7999 \exp( - 4.s)/(4.s) - 2134 \exp( - 2.5s) / (2.5s)
\label{seqn13}
\end{equation}   
\noindent
This interaction is based upon a realistic G-matrix. Since the G-matrix was constructed in an oscillator representation, it is effectively an average over a range of nuclear densities and therefore the M3Y has no explicit density dependence. For the same reason there is also an average over energy and the M3Y has no explicit energy dependence either. The only energy dependent effects that arises from its use is a rather weak one contained in an approximate treatment of single-nucleon knock-on exchange. The success of the extensive analysis \cite{r10,r11} indicates that these two averages are adequate for the real part of the optical potential for heavy ions at energies per nucleon of $< 20MeV$. However, it is important to consider the density and energy dependence explicitly for the analysis of $\alpha$-particle scattering at higher energies ($>100 MeV$) where the effects of a nuclear rainbow are seen and hence the scattering becomes sensitive to the potential at small radii. Such cases were studied introducing suitable and semirealistic explicit density dependence \cite{r12,r13} into the M3Y interaction which was then called the DDM3Y and was very successful for interpreting consistently the high energy elastic $\alpha$ scattering data.

      The entire calculations have been redone again using the density dependent M3Y effective interaction (DDM3Y) supplemented by a zero-range pseudo potential, representing the single nucleon exchange term \cite{r13}. In DDM3Y the effective nucleon-nucleon interaction $v(s)$ is assumed to be density and energy dependent and therefore becomes functions of density and energy and is generally written as 

\begin{equation}
  v(s,\rho_1,\rho_2,E) = t^{M3Y}(s,E)g(\rho_1,\rho_2,E)
\label{seqn14}
\end{equation}   
\noindent
where $t^{M3Y}$ is the same M3Y interaction given by eqn.(13) but supplemented by a zero range pseudo-potential 

\begin{equation}
  t^{M3Y} = 7999 \exp( - 4.s) / (4.s) - 2134 \exp( - 2.5s) / (2.5s) + J_{00}(E) \delta(s)
\label{seqn15}
\end{equation}   
\noindent
where the zero-range pseudo-potential representing the single-nucleon exchange term is given by

\begin{equation}
 J_{00}(E) = -276 (1 - 0.005E / A_\alpha ) (MeV.fm^3)
\label{seqn16}
\end{equation}   
\noindent
and the density dependent part has been taken to be \cite{r13}

\begin{equation}
 g(\rho_1, \rho_2, E) = C (1 - \beta(E)\rho_1^{2/3}) (1 - \beta(E)\rho_2^{2/3})
\label{seqn17}
\end{equation}   
\noindent
which takes care of the higher order exchange effects and the Pauli blocking effects.

      The energy E appearing in the above equations is the energy measured in the centre of mass of the $\alpha$ - daughter nucleus system and for the $\alpha$ decay process it is equal to the released energy Q. Since the released energies involved in the $\alpha$ decay processes are very small compared to the energies involved in high energy $\alpha$ scattering, $\beta(E)$ has been considered as a constant and independent of energy. The zero-range pseudo-potential $J_{00}(E)$ is also practically independent of energy for the $\alpha$ decay processes and can be taken as $-276 MeV.fm^3$. It is important to mention here that use of the nuclear interaction potential for exotic cluster decays obtained by double folding an effective nucleon-nucleon interaction with their respective densities was suggested in reference \cite{r14}. After the WKB action integral $K$, given by eqn.(10) has been evaluated, the half lives of the $\alpha$ decays have been calculated using eqn.(9) and eqn.(12).  

\section{RESULTS AND DISCUSSIONS}
\label{section4}

      For the present illustrative calculations, the same set of experimental data of reference \cite{r15} along with the rest from reference \cite{r16} for the $\alpha$ decay half lives have been chosen for comparison with the present theoretical calculations for even-even parent nuclei for which the experimental ground state masses for the parent and daughter nuclei are available. This set was selected because there is no uncertainty in the determination of the released energy Q (given by eqn.(8)) which is one of the crucial quantity for quantitative prediction of decay half lives and all the the parent and daughter nuclei have zero spins and positive parities. For the calculations using DDM3Y with zero range exchange also, the value of  the normalization C appearing in eqn.(17) has been fixed at 1.0. The density dependent parameter $\beta(E)$, which is supposed to be dependent on energy, has been kept constant and independent of Q. Optimum fit to the data has been obtained using $\beta(E)$=1.6. 

\begin{figure}[h]
\eject\centerline{\epsfig{file=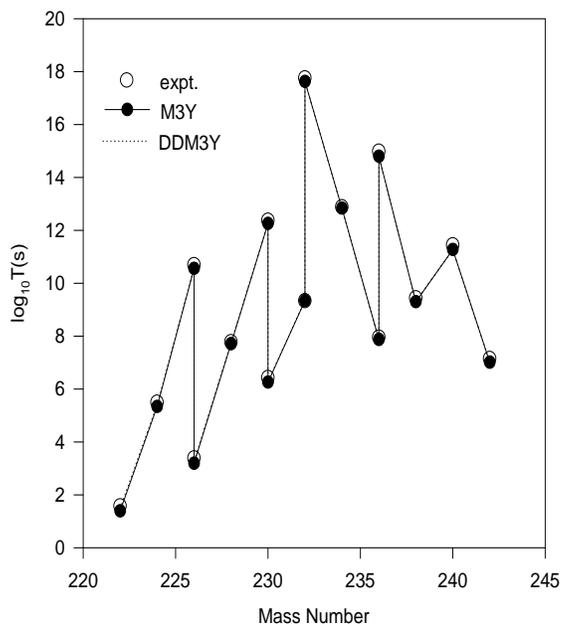,height=15cm,width=10cm}}
\caption
{Logarithmic $\alpha$-radioactivity half-lives plotted against the parent mass number. The experimental data have been shown by the open circles. The solid circles represent the results of present calculations, using M3Y effective interaction, for the corresponding experimental data . The continuous line connects these calculated values. The dotted line (almost indistinguishable from the continuous line) represent the same calculations using DDM3Y with a zero-range pseudo-potential for the nucleon-nucleon effective interaction. For both the theoretical calculations  spherical charge distributions for the Coulmb interaction has been assumed. }
\label{fig1}
\end{figure}

      In Fig.~\ref{fig1} the experimental data for logarithmic $\alpha$ decay half lives \cite{r15,r16} have been plotted against the mass numbers of parent nuclei along with the results of the present calculations for zero angular momentum of the fragments. In the figure the open circles depict the experimental data while the continuous line with solid circles represents the corresponding calculations of the present model using the M3Y effective interaction and the dotted line represents the present calculations using DDM3Y effective interaction supplemented by a zero-range pseudo-potential. The decay modes and the experimental values for their half lives have been presented in Table 1. The corresponding results of the present calculations of superasymmetric fission model with microscopic potentials are also presented along with the results of the liquid drop model (LDM) calculations of 2001 so as to facilitate the comparison of the results of the phenomenological calculations with the present one. The exact Q values calculated by eqn.(8) using the experimental ground state masses \cite{r17} and used by the present calculations have also been presented. 

      The results of the present calculations with M3Y or DDM3Y with pseudopotential have been found to predict the general trend of experimental data very well. The quantitative agreement with experimental data is excellent. The degree of reliability of the  present estimates for the $\alpha$ decay lifetimes are better than the liquid drop description \cite{r15} for most of the cases. The recent result of the preformed cluster model \cite{r18} calculation for the $\alpha$ decay half-life of $^{242} Cm$ is also much worse compared to that obtained from present calculations.  The degree of reliability of the  present estimates for the $\alpha$ decay lifetimes is better than the recent results of the LDM calculations \cite{r19} presented in the table below.  

\begin{table}
\caption{Comparison between Calculated $\alpha$-decay Half-Lives using Point (PCD) and Spherical charge distributions (SCD) respectively for the Coulomb interaction and using effective interactions of M3Y and DDM3Y with zero-range pseudo-potential for the nuclear interaction.}
\begin{tabular}{ccccccc}
Parent &       &LDM(01)&Present  & M3Y(DDM3Y+ &M3Y(DDM3Y+ & Expt.      \\ 
          &             &        &     &pseudopotential)&pseudopotential)&                \\ 
 & & & &SCD&PCD&                \\ \hline
$Z$&$A$&$log_{10}T(s)$&$Q(MeV)$&$log_{10}T(s)$&$log_{10}T(s)$&$log_{10}T(s)$ \\ \hline

  88 &222  &1.72  &6.68 &1.39(1.50)&1.43(1.52)      &      1.58     \\ 
  88 &224  &5.74   &5.79 &5.34(5.41)&5.39(5.44)          &  5.50       \\ 
  88 &226  &10.98 &4.87 &10.56(10.60)&10.62(10.63)  &         10.703    \\  
  90 &228  &8.07   &5.53 &7.71(7.75)&7.77(7.78)&       7.781    \\
  90 &230  &12.75 &4.78 &12.26(12.28)&12.34(12.31)&    12.376     \\ 
  92 &232  &9.69   &5.42 &9.34(9.36)&9.42(9.39)&  9.337     \\
  92 &234  &13.21 &4.86 &12.84(12.83)&12.91(12.87)    &        12.889    \\   
  94 &236  &8.01&5.87 &7.87(7.87)&7.96(7.91)&      7.954      \\ 
  94 &238  &9.54   &5.60 &9.30(9.31)&9.39(9.35)   &   9.4423    \\ 
  96 &242  &7.11   &6.22 &7.01(7.02)&7.11(7.06)&7.1485    \\ 
  90 &226  &3.49   &6.46 &3.19(3.27) &3.24(3.30)   & 3.39 $^a$ \\
  90 &232  &18.18 &4.08 &17.63(17.62)&17.72(17.66) & 17.76 $^a$ \\
  92 &230  &6.54   &6.00 &6.26(6.30) & 6.32(6.33)& 6.43 $ ^a$ \\
  92 &236  &15.23 &4.58 &14.80(14.78)&14.89(14.82)& 14.99 $^a$ \\
  94 &240  &13.39 &5.26 &11.28(11.27)&11.38(11.30)& 11.45 $^a$ \\
\end{tabular} 
$(a)$ Taken from reference \cite{r16}.
\end{table}

      Above results show that the differences in the $\alpha$-decay lifetimes obtained using the spherical charge distributions or point charge distributions for calculating the Coulomb interaction are small. Introducing the density-dependence and the zero-range pseudopotential do not alter the general trend of the results and also do not improve the quality of fit to the experimental data significantly. Use of the M3Y effective interaction alone for calculating the double folding nuclear interaction potentials is sufficient to provide reasonable estimates for the $\alpha$-decay lifetimes.     

\section{CONCLUSIONS}
\label{section5}

      The half lives for $\alpha$-radioactivity have been analyzed with microscopic nuclear potentials obtained by the double folding pocedure using M3Y and DDM3Y effective interactions respectively. This procedure of obtaining nuclear interaction potentials is based on profound theoretical basis. The results of the present calculations using M3Y or DDM3Y supplemented by a pseudo-potential are in good agreement over a wide range of experimental data.  Present calculations show that the differences in the results of the $\alpha$ decay lifetimes obtained using the spherical charge distributions or point charge distributions for calculating the Coulomb interaction are small. Use of the density-dependence and the zero-range pseudopotential also neither alters the general trend of the results nor improves significantly the quality of fit to the experimental data. Refinements such as introduction of dissipation while tunneling through the barrier or incorporating the dynamic shape deformations in the density distributions of the clusters may further improve results. It is worthwhile to mention that using the realistic microscopic nuclear interaction potentials, the results obtained for the $\alpha$ radioactive decay lifetimes are noteworthy and are comparable to the best available theoretical calculations. Such calculations may be extended to provide reasonable estimates of the lifetimes of nuclear decays by $\alpha$ emissions for the entire domain of exotic nuclei.

\end{document}